\documentclass[10pts]{article}\usepackage{graphicx}

\begin{document}


\date{}

\title{Setting the Agenda: Different strategies of a Mass Media
in a model of cultural dissemination}

\author{Sebasti\'an Pinto$^1$, Pablo Balenzuela$^{1,2}$, Claudio O. Dorso$^{1,2}$}

\maketitle

\thanks{\small $^1$ Departamento de F\'isica, Facultad de Ciencias Exactas y Naturales, Universidad de Buenos Aires, Av. Cantilo s/n, Pabell\'on 1, Ciudad Universitaria, 1428, Buenos Aires, Argentina.}\\
\thanks{\indent \small $^2$ Instituto de F\'isica de Buenos Aires (IFIBA), CONICET, Av. Cantilo s/n, Pabell\'on 1, Ciudad Universitaria, 1428, Buenos Aires, Argentina.}\\

\begin{abstract}

Day by day, people exchange opinions about a given new
with relatives, friends, and coworkers.
In most cases, they get informed about a given issue by 
reading newspapers, listening to the radio, or 
watching TV, i.e., through a Mass Media ({\it MM}).
However, the importance of a given new 
can be stimulated by the Media 
by assigning newspaper's pages
or time in TV programs. In this sense, we say that 
the Media has the power to ``set the agenda", i.e., 
it decides which new is 
important and which is not.
On the other hand, the Media 
can know people's concerns through,
for instance, websites or
blogs where they express their opinions,
and then it can use this information 
in order to be more appealing to an
increasing number of people.
In this work, we study different scenarios in 
an agent-based model of cultural dissemination, in which 
a given Mass Media 
has a specific purpose: To set a particular topic
of discussion and impose its point of view to as many social agents as it can.
We model this by making the Media 
has a fixed feature, representing its point of view
in the topic of discussion, 
while it tries to attract new consumers, 
by taking advantage of feedback mechanisms,
represented by adaptive features.
We explore different strategies that the Media can 
adopt in order to increase the affinity with potential consumers
and then the probability to be successful in imposing this particular 
topic.

\end{abstract}


\section{Introduction}\label{sec:Introduction}

 
Many times, people get involved in discussions about 
certain issues that don't arise only from their own daily 
experiences, in the sense that these seem to behoove
the social group to which they belong, such as discussions
about their country's macro economy, regional elections, etc.
Due to the complexity and variety of those issues, and
in many cases the remoteness with the situation,
people resort to Mass Media in order to get informed
about these ones and to know the opinion
of specialists in these topics.
People become interested in these issues because 
the Media is supposed to reflect the interests and
concerns of their social environment.\par

Following Giddens (\cite{Giddens}), we find several	
theoretical approaches to the role of a Mass Media 
in the field of sociology. A Media seen as
social stabilizer, which keeps and reflects the dominant
culture, is the basis of the functionalism theory.
As Giddens says, several reasons lead sociologists to
move away from this approach: One of them 
is that the functions mentioned above appear wholly
positive. In contrast to functionalism, the conflict theory 
sees the Media as a less benign force within society:
It is a powerful agent whose ideology 
justifies or legitimizes the interests of the 
owner group of the Media.
The ideology of a Media can be explicit, as
for instance, in the editorial line of many newspapers,
but in most cases it's implicit in the TV time 
or newspaper's pages that the Media spends to discuss
a particular issue. 
The imposing of a topic in public opinion is what
is called ``to set the agenda", widely analyzed 
by Mccombs (\cite{Mccombs} \cite{Mccombs1972}). As it can be 
read in \cite{Mccombs1972}, 
\emph{``(the press) may not be successful much of the time 
in telling people what to think, but it is stunningly
successful in telling its readers what to think about"}.
However, during the coverage of a given issue,
the Media can suggest its point of view to the audience.\par

We analyze this idea in an agent-based model of cultural 
dissemination (the Axelrod's model \cite{Axelrod}, see section \ref{sec:TheModel}),
where each individual is characterized by a set of 
features representing 
its cultural profile, who interact proportionally to their 
degree of similarity 
(Homophily).
Specifically, in this work, we analyze the case where a Mass Media 
has a given purpose:  
It is interested in ``setting the agenda", i.e. 
make the largest amount of agents discuss about a given topic,
as for instance, a particular policy issue, and impose its point of view.
To pursue this goal, in our model
the {\it MM}  is able to modify  the topic of discussion in 
each feature following different
strategies. This acts as a 
feedback mechanism in order to be more appealing 
to the majority of the agents and increase
the probability of interaction 
with them, in line with the reported in  \cite{Wood} 
where individuals sharing common attributes tend to 
be more similar.\par
In this work, we interpret each agent's value of
a given feature
as the main interest in this particular topic, as for instance,
its favorite sport or its opinion about a policy issue.
The Axelrod's model is very well suited to study 
the influence 
of a {\it MM} over a given population because each feature 
could be naturally 
interpreted as the section of a given newspaper. For instance 
the New York Times 
present the following sections: World, U.S., Politics, N.Y, Business, 
Opinion, 
Technology, Sports, Health, Science, Arts, Fashion and Style, and 
Food.\par 

\subsection{Previous works}
Previous works in this topic basically follow two approaches:
a fixed Mass Media, whose cultural state is constant in
time and represents a Media who
has no feedback with the population,
and a fully adaptive Mass Media,
which varies its cultural state adopting the
most popular trait in each feature.\par

From a social point of view, a constant Mass Media 
represents a Media who impose the topic of discussion in all features 
regardless the society concerning.
From the physical point of view, it acts as an external constant 
vector field who drives the states of the agents.
This modeling approach was followed in 
\cite{GonzalezAvella2005} 
and \cite{Mazzitello2007}. In the first one, the authors 
studied the combined dependence of the stationary states 
with the number 
of traits per feature ($Q$) and the probability of interaction 
with the {\it MM} ($B$). 
They counter-intuitively found that the 
Mass Media induces cultural diversity when the interaction 
parameter $B$ is above certain threshold. 
In the second work, 
the combined effects of a fixed {\it MM} and a cultural drift 
(modeled as random perturbations) was analyzed. 
They also 
included an extra feature which make the interaction 
between the {\it MM} and the agents always possible.
An interesting twist was followed in \cite{ARodriguez2009} 
where the Mass Media is characterized by two parameters: 
a non-null overlap with all agents and a confidence value 
of its information. The first parameter is related to the 
concept of ``propaganda", by which the {\it MM} can
interact with all agents, included those cases where there
is no cultural similarity. The second 
parameter is intended to model the level of credibility 
of a $MM$ which, according to the authors, 
is directly related to its level of influence. 
A similar approach was followed in \cite{ARodriguez2010},
where the authors incorporate the influence of the 
Mass Media as a non-pairwise interactions among agents, 
following the proposal of \cite{FlacheMacy2011} 
for the Axelrod's Model.\par

The other approach includes
 feedback processes between the Mass Media 
and the social community. 
In all the cases, the Media adopts the 
most popular point of view in each 
feature. From a social point of view, 
this modeling approach is 
closer to the functionalism theory 
described in \cite{Giddens}, where 
the Media is supposed to reflect the dominant
culture of a society.
On the other hand, from the physical point of view,
the Media only catalyzes the dynamics toward 
consensus of the population, i.e., the Media
doesn't induce any particular state.
This problem was initially faced in \cite{Shibanai2001} 
following an 
Axelrod's model, where two different 
variants were proposed: A global field 
where each feature of the {\it MM} adopted 
the most popular point
of view of the population and a filter model, where 
the feedback
is indirectly modeled in the interaction between agents. 
In \cite{GonzalezAvella2006}, the authors proposed three 
different 
ways in which the Mass Media  could be modeled: as an 
external field (as a fixed Mass Media), 
as a global field (where the {\it MM} adopts the most 
popular point 
of view of each feature for all of them, making it time dependent 
but uniformly distributed in space) and as a local field 
(where the field adopted the most popular point of 
view among 
an agent's neighbors, i. e. it is non-constant in space 
and time). 
In \cite{GonzalezAvella2007}, the authors also systematically 
investigated the indirect feedback mechanism as proposed 
in \cite{Shibanai2001}. It is important to remark that, 
in all cases, the feedback between the {\it MM} 
and the population was present in all the features.\par

Many other studies have been made in the 
context of the Axelrod's model.
The role of the social contact network in the dynamics 
with a fixed {\it MM} was also 
investigated in \cite{Candia2008}, where the effects 
of intra and inter-links of a social network with 
community structure was analyzed. 
In \cite{LatoraMoreno2010}, the microscopic 
dynamics toward equilibrium was analyzed 
when the underlying network is scale-free 
in its degree distribution.
In the same modeling scenario, a model of 
cross-cultural interaction through Mass Media 
interactions was investigated in \cite{GonzalezAvella2012}, 
where two (fully adaptive) Mass Media act over two different 
interconnected populations. In this model, 
one of the  Mass Media reflects the dominant 
cultural state of a given population and 
influence the other one. 
The study of social interactions and the presence 
of a Mass Media was also explored in the context 
of other models, as for instance, 
the Deffuant's model (\cite{Pineda2015}), 
the voter model (\cite{Masuda2015}), 
and the Sznadj's model 
(\cite{Zhao2015},\cite{Crokidakis2012}).\par

\subsection{Our contribution}
As was mentioned above, we consider the Axelrod's 
model (see section \ref{sec:TheModel}) as the 
best candidate to study the social influence of a Mass Media,
because of the natural interpretation of Media's cultural state
as the sections of a given newspaper.
We also mentioned that the previous works 
follow basically two approaches: a fixed 
and a fully-adaptive Mass Media.
While a fixed Mass Media is an 
oversimplification of its actual role in a society, 
mainly because of the absence of a feedback mechanism 
between the Media and the population,
a fully adaptive implementation suites very well
to the functionalism theory of Media's influence, 
but, as Giddens says (\cite{Giddens}), this theory have fallen 
into decline in the recent decades, because 
it presents the Media in a very naive way, as 
an external agent without ideology or purposes, who 
only reflects the dominant culture.

In this work, we model the Mass Media as an external agent 
with some features fixed and the rest adaptive.
Despite the apparent little difference 
between our model and the approaches mentioned
above, we consider that our interpretation and representation 
of the Media fits better within the conflict 
theory of Media's influence (\cite{Giddens}) and within 
the works of Mccombs (\cite{Mccombs} \cite{Mccombs1972}). 
Here, the Media influences 
the population with a given purpose: 
To put an special topic to be discussed 
by public opinion, i.e., to set the ``agenda"
on a particular feature, and impose its point of view. 
From now on we will refer to this peculiar
value of the selected feature as the Mass Media's topic
({\it MMT}). 
Simultaneously, it will try to adapt the rest 
of its features in order to attract a great number 
of consumers.
We will explore two different strategies in order to do that:
A conservative one, where
it looks for increasing the number of followers,
from a well established group of them, and an aggressive one,
in which the Mass Media targets all those individuals which
have not attached yet.
From now we call {\it Followers} to those agents 
who adopt the {\it MMT}.
We will explore the different collective dynamics which emerges
with these strategies. We compare the results with the
case where the {\it MM} doesn't follow any strategy, i.e.,
it is constant in time.\par
The work is organized as follows: 
In section \ref{sec:TheModel} we describe the model 
that we implemented for our numerical simulations, 
describing the different strategies 
that the Mass Media can adopt, 
and the definition of the observables analyzed. 
In section \ref{sec:Results} 
we show the main results concerning as well as the 
equilibrium properties and the dynamics 
towards the equilibrium of different 
Mass Media's strategies. 
In particular, we will be interested in
the total number of followers as a 
function of time and their self-similarity and
similarity with the Mass Media.
In section \ref{sec:Conclusions} we present the 
conclusions of the work.\par

\section{The Model} \label{sec:TheModel}

In this work, there are two main actors, both
described within the Axelrod's model: On the one hand
we have a population of agents which interact
amongst them and, on the other, the Mass Media, 
which interacts with all the members of the
population.

\subsection{The Axelrod Model}

The Axelrod's model \cite{Axelrod} 
is an agent based model 
which assumes that the cultural state of 
each individual can be described in 
terms of a set of attributes such as 
political orientation, religion, 
sports preferences, etc. 
The interaction mechanism between 
agents is pairwise based and 
rests on two fundamental hypothesis:
\begin{itemize}
\item Homophily: the probability 
of interaction between two 
individuals is proportional 
to their cultural similarity.
\item Social Influence: after each interaction, 
the agents become more similar. (see section \ref{sec:Dynamics})
\end{itemize} 
The success of the original model is due 
to the emergence of a non-intuitive 
stationary collective behavior: a 
transition between a monocultural global state, 
in which all the agents are identical, 
and a state of cultural diversity, 
characterized by the coexistence of 
regions with different cultural states.\par

\subsubsection{The Population}

We implement the Axelrod's model 
with $N$ agents placed in the nodes of a 
two-dimensional grid, 
with rigid walls, i. e.,
the system is finite.
Following \cite{Axelrod}, 
the cultural state of each agent 
can be represented by a vector 
$v=(v_1,v_2,....,v_F)$ where $F$ 
stands for the number of features. 
Each component $v_i$ is a 
nominal variable corresponding 
to a certain cultural feature 
and can adopt $Q$ different values, 
representing 
the different traits in a specific feature. 
We interpret the value of
a given feature
as the main interest in this particular topic,
and this interpretation is 
analogous that we give to the Mass Media's 
state, as we describe below.

\subsubsection{The Mass Media}

The Mass Media is modeled as an 
external agent, 
with the same number of features 
and traits than the agents, 
which, in principle, 
can interact with all of them with probability $B$.
In this work, the Mass Media's state
represents the sections of a newspaper,
and each feature's value, the main theme 
covered en each topic.
The {\it MM}'s state 
has a fixed value in the first component, i.e., 
$v_1^{MM}=1$, and represents the {\it MMT}
defined above.
The other features
fluctuate in time according 
to different strategies, 
as we detail bellow. In what follows, 
we will call {\it Followers} to those agents 
in the population who share the {\it MMT},
i.e., agents with $v_1 = 1$, 
independently of the other
features' values.
On the other hand, the {\it Non-Followers} 
are those agents with $v_1 \neq 1$. 
In order to increase the interaction 
probability with the majority of 
the population and potentially 
increase the amount of {\it Followers}, 
the Mass Media can change the other 
features according to one of 
the following strategies: 

\begin{itemize}
\item The Followers Strategy ({\it FS}): 
In the non-fixed components of its cultural vector 
($v_2-v_F$), the Mass Media adopts, 
at each time step, the most abundant 
value among those agents who share 
the {\it MMT}, i.e., 
{\it Followers} agents. 
This is a conservative strategy and 
its main goal is to increase 
the amount of $Follower$ 
from a well consolidated crew.
\item The Non-Followers Strategy ({\it NFS}): 
In the non-fixed components, 
the Mass Media adopts the most 
abundant value among those agents 
who don't share the 
{\it MMT}, 
i.e., {\it Non-Followers} agents, 
in order to maximize the probability of 
interaction with them, and convince 
them rapidly. In opposite of {\it FS}, 
this is an aggressive or conqueror strategy. 
\end{itemize}
In all cases, we compare our results 
with the case of a Fixed Mass Media 
({\it FMM}), where all the features of 
the Mass Media remains constant in time, 
as it was analyzed 
in \cite{GonzalezAvella2005}. \par

\subsubsection{Dynamics} \label{sec:Dynamics}
The dynamics  of the model is the following:
\begin{itemize}
\item Select one element $i$ from the lattice.
\item Select another element $j$, 
which with probability $B$, $j=MM$, 
and with probability $(1-B)$, $j$ 
is one of the nearest neighbors of $i$, 
selected at random.
\item The probability of interaction between 
agents $i$ and $j$, $P_{i,j}$, 
is given by the fraction of shared features, 
$P_{i,j}=\frac{1}{N}\sum_{k=1}^F \delta_{v_{k}^{i},v_{k}^{j}}$. 
We will refer to this probability of 
interaction as the {\it overlap} 
between agents $i$ and $j$.
\item If  $P_{i,j} \ne 0$ and $P_{i,j} \ne 1$, 
then agent $i$ picks at random a feature 
$v_{k}^{i}$ and adopts the corresponding 
trait of the agent $j$, $v_{k}^{j}$ 
(but it doesn't change immediately, see the next step).
\item We repeat this task for all 
the agents in the system, 
updating the changes synchronously. 
This is what we call a time step.
\item After a time step, the Mass Media's state 
is updated according to the current strategy.
\end{itemize}

\subsection{Observables}

In order to study the behavior of the system 
according to the different strategies 
quoted above, 
we define the following observables: 
\begin{itemize}

\item \underline{Fraction of {\it Followers} ($F/N$)}: 
It's the fraction of agents who 
share the first feature's value
with the Mass Media. The fraction of 
{\it Followers} is the main  observable
in order to evaluate
the effectiveness of each strategy. 
As it can be noticed this quantity
can only be defined in this modeling approach
(i.e., when we have only one feature fixed).

\item \underline{The normalized size 
of the biggest fragment ($S_{max}/N$)}: 
It represents the largest group of 
connected agents who share
the first feature's value. ($S_{max}/N$) 
is an standard quantity in order to study 
collective properties in the Axelrod's model. 
The two classical stationary solutions, 
consensus and cultural diversity, 
can be easily identified by studying 
the behavior of ($S_{max}/N$) 
as function of the system's parameters.

\end{itemize}

It is important to remark 
that being a Follower agent 
only implies that it shares
the first feature's value 
with the Media, independently
of the others. It's 
interpreted as it adopts the {\it MMT}, 
but maybe it's not interested
in the other Media sections. 
On the other hand, 
it is important to stress at this point 
that, given an ensemble of realizations, 
the features' values different to the one 
corresponding to the {\it MMT} will be 
homogeneously distributed amongst 
all the elements of the space 
of realizations. On the other 
hand, depending of the values 
of $B$ and $Q$, the feature corresponding 
to the {\it MMT} will end to attain 
the value ``pushed" 
by the Media.\par

In order to have a map of all 
stationary solutions of the model,
we will plot a Mass Phase Diagram 
({\it MPD}), where we calculate ($F/N$) 
as a function of $B$ and $Q$. 
With this, we can explore how
effective is the Mass Media to convince
as many agents as it can.
On the other hand, 
we will plot a Maximum Cluster Phase Diagram 
({\it MCPD}) where we calculate 
($S_{max}/N$) as a function 
of $B$ and $Q$ (\cite{Cosenza2010}). 
It takes into account cluster properties, and
it's not necessarily a cluster composed by 
{\it Followers}. Both phase diagrams have been made 
for all the strategies defined above.\par
We are  also interested in studying 
the average similarity among the {\it Followers}, 
so we define the following quantities:
\begin{itemize}
\item The mean homophily between 
the Mass Media and the {\it Followers}:
\begin{equation}
H_{MM}(t)=\frac{1}{N'}\sum_{i}^{N'} 
(\sum_{k=1}^{F}\delta_{v_{k}^{i},v_{k}^{MM}})
\end{equation}
 where the first sum is over the $N'$ 
{\it Followers} and the second one 
over the amount of features. 
This quantity takes into account the average 
similarity between the 
{\it Followers} and the Mass Media.
\item The mean homophily among the {\it Followers}:
\begin{equation}
H_F(t)=\frac{1}{M'}\sum_{i<j}^{M'} 
(\sum_{k=1}^{F}\delta_{v_{k}^{i},v_{k}^{j}})
\end{equation}
where $M'$ stands for all the pairs of {\it Followers} 
that can be formed, and the first sum is over 
all agents ($i$ and $j$) who are {\it Followers}. 
This quantity expresses 
the average similarity among {\it Followers}.
\end{itemize}

As the states of the agents vary with time, 
these observables are time-dependent 
and will bring useful information about 
the dynamical behavior of the system.\par

\section{Results} \label{sec:Results}

We performed numerical simulations using 
a two-dimensional finite grid of $50 \times 50$ 
nodes (total number of nodes $N=2500$). 
In each node $i$, ($i=1,...,N$), an agent 
with a given cultural vector $v_1,....,v_F$ 
is placed. The number of features is 
$F=10$ and represents the typical 
number of sections of a  newspaper 
(for instance, in the web edition of 
New York Times, the main newspaper's 
sections are thirteen: World, 
U.S., Politics, N.Y,. Business, Opinion, 
Tech, Science, Health, Sports, Arts, 
Fashion and Style, and Food. 
In the international edition 
of ``El Pa\'is" from Spain there 
are ten sections.).\par

\subsection{Equilibrium Properties}

In this section we study the characteristics 
of the stationary states. In these states 
the overlap between any pair of agents 
(including the Mass Media) is zero or one.
This implies that a {\it Follower} agent 
finishes to share all
features' values with the {\it MM}.
In this model, the system always reaches a 
stationary state.\par
\begin{figure}
\centering
\includegraphics[width = \textwidth]{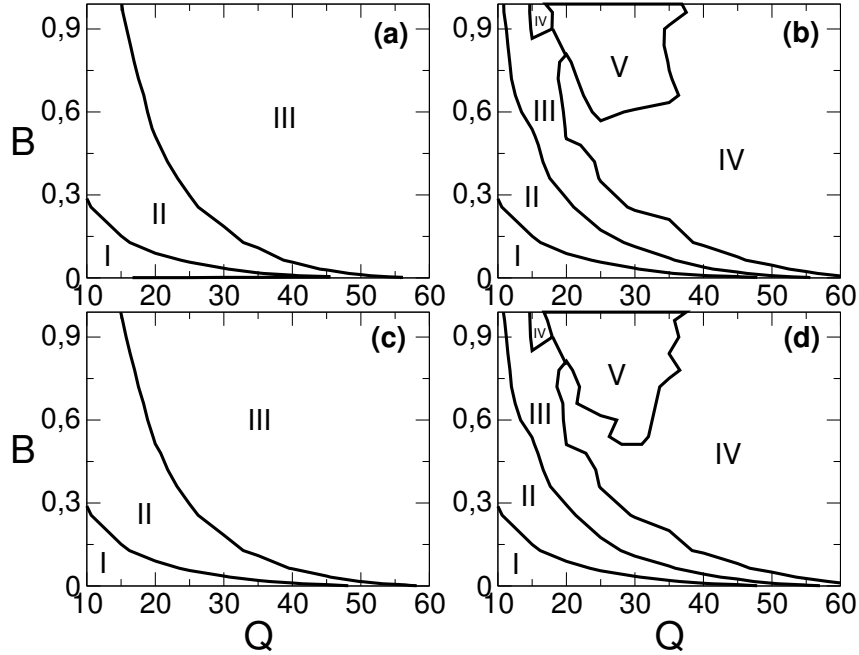}
\caption{\textbf{Phase Diagrams}. 
Mass Phase Diagram 
(Left Panels, (a) and (c))
and
Maximum Cluster Phase Diagram 
(Right Panels, (b) and (d)) 
for a Fixed Mass Media ({\it FMM}, 
top panels, (a) and (b)) and 
Followers Strategy ({\it FS}, 
down panels, (c) and (d)). 
Five regions can be identified 
according the degree of 
dominance of a given state 
which are detailed in the main text. 
The phase diagrams corresponding 
to the Non-Followers Strategy ({\it NFS}) 
are not shown because there is 
only one solution in the range 
of analyzed parameters: 
consensus with the {\it MM} (Region {\it I}).}
\label{Fig1}
\end{figure}

In Fig.\ref{Fig1}, we plot the Mass Phase 
Diagram ({\it MPD}) and the 
Maximum Cluster Phase 
Diagram ({\it MCPD}) for the Followers 
Strategy and Fixed Mass Media ({\it FS} 
and {\it FMM}), respectively. 
Three regions can be identified in the {\it MPD}
corresponding to 
different kind of stationary solutions:

\begin{enumerate}
\item[I] \underline{Consensus identical to the {\it MM}}: 
above the $90\%$ of the agents have 
the same cultural state than the {\it MM}. 
This region points out the hegemony of the {\it MM}.
\item[II] \underline{Absolute Dominance of {\it MM}}: 
this region is characterized by a dominant 
mass, identical to the {\it MM}, 
whose size is above the $50\%$ and bellow 
the $90\%$ of the population.
\item[III] \underline{Relative Dominance of {\it MM}}: 
this region is characterized by a dominant 
mass, identical to the {\it MM}, 
whose size is above the $10\%$ and 
bellow the $50\%$ of the population.
\end{enumerate}

These regions can also be found in the {\it MCPD},
if we replace the term {\it mass} for {\it cluster}, i. e., 
we find a maximum cluster whose relative 
size is above the 
$90\%$ (region {\it I}), between the $90\%$ and the $50\%$
(region {\it II}), and between the $50\%$ and the $10\%$ of the population
(region {\it III}).
In all these cases, the maximum cluster corresponds to a Mass
Media's state.
However in the {\it MCPD}, two more regions can be
identified:

\begin{enumerate}
\item[IV] \underline{Fragmentation}: 
there is no dominant clusters 
of agents. The size of the biggest 
cluster is smaller than 
the $10\%$ of the  population.
\item[V] \underline{Local Relative Dominance}: 
this region is characterized by a dominant 
cluster with a different state respect 
to the Mass Media's one, 
whose size is above the $10\%$ and 
bellow the $25\%$ of the  population. 
\end{enumerate}

For the values of B and Q explored in the phase
diagrams, the fraction of $Followers$ always 
exceeded the $10\%$ of the population,
although it does not necessarily form an unique cluster:
This is why the region {\it IV} in the {\it MPCD} 
and the region {\it III}
in the {\it MPD} can coexist. 
On the other hand, 
it is important to note the presence 
of a region dominated by a cluster 
whose state is different to the 
Mass Media's one that we call region {\it V}.
This region was 
reported in \cite{Cosenza2010} for a 
Fixed Mass Media,
and it acquires more relevance
in networks
with long-range interactions.
We find this region also for 
the Followers strategy's {\it MCPD}.\par
An important observation of Fig.\ref{Fig1} 
is the absence 
of a phase diagram corresponding to the 
Non-Followers Strategy ({\it NFS}). 
We haven't plot it because, with this strategy, 
there is only one stationary state: 
consensus similar to the Mass Media. 
This is a fingerprint of this strategy: 
it is able to produce consensus 
for any values of $B$ and $Q$.
In this strategy, the Mass Media adapts its 
non-fixed features in order to maximize 
the interaction probability with 
those agents who don't share 
the {\it MMT}. 
Therefore, the Media is able to make 
all agents become {\it Followers}. 
Once this task is completed, the Mass Media 
produces no further changes in its state. 
The remaining dynamics corresponds to 
interactions between agents in order to 
reach total consensus according the Axelrod 
model's dynamics. Even though this strategy 
shows only one equilibrium solution, 
its dynamical behavior shows 
a dependence with the parameters of the system, 
as we will show in the next section. \par

On the other hand, for a Fixed Mass Media 
({\it FMM}) and the Followers Strategy ({\it FS}), 
both phase diagrams are qualitatively similar: 
the dominance of the Mass Media state is 
absolute for low $Q$ and $B$ (left-bottom corner) 
and it losses preponderance when $Q$ and 
$B$ increases. In the top-right corner of 
the plots ($Q\simeq 60$ and $B \simeq 0.9$) 
between the $10\%$ and the $50\%$ of the agents 
share the {\it MMT}, but
there is no cluster in the 
system bigger than the $10\%$ 
of the lattice's population.
Also, for both Fixed Mass 
Media and Followers Strategy's {\it MPCD}, 
the region {\it V} is present,  i. e., the 
maximum cluster is orthogonal to the Mass Media, 
but its size doesn't exceed 
the total amount of {\it Followers} present in the system.\par 

In what follows we will analyze which are the 
main characteristics and differences between 
the collective dynamical behavior of the 
population for the different strategies 
followed by the Mass Media.\par

\subsection{Dynamical properties of collective states 
for different strategies}

In the analysis of equilibrium states, we have seen that a 
Fixed Mass Media and the Followers Strategy show 
similar phase diagrams, while for the Non-Followers 
Strategy the system evolves to a consensus with 
the Mass Media for all values of $B$ and $Q$. 
Given these known equilibrium properties, 
the questions we would like to face in this section are two: 
\begin{enumerate}
\item How is the dynamics toward equilibrium of 
the system and the Mass Media for each strategy?
\item Do the Mass Media's followers form an 
homogeneous or an heterogeneous cultural group?
\end{enumerate}

With this in mind, we analyze the temporal evolution 
of the system for a case of low probability interaction 
with the Mass Media ($B=0.01$) and two different 
values of $Q$: $Q=20$ and $Q=60$.
For $Q=20$, all the strategies reach the consensus of 
the whole population, meanwhile, for $Q=60$, 
only {\it NFS} does. 
We analyze the collective behavior of 
the population in both cases in terms 
of ($F/N$), $H_{MM}$, and $H_{F}$.\par

\begin{figure}
\includegraphics[width = \textwidth]{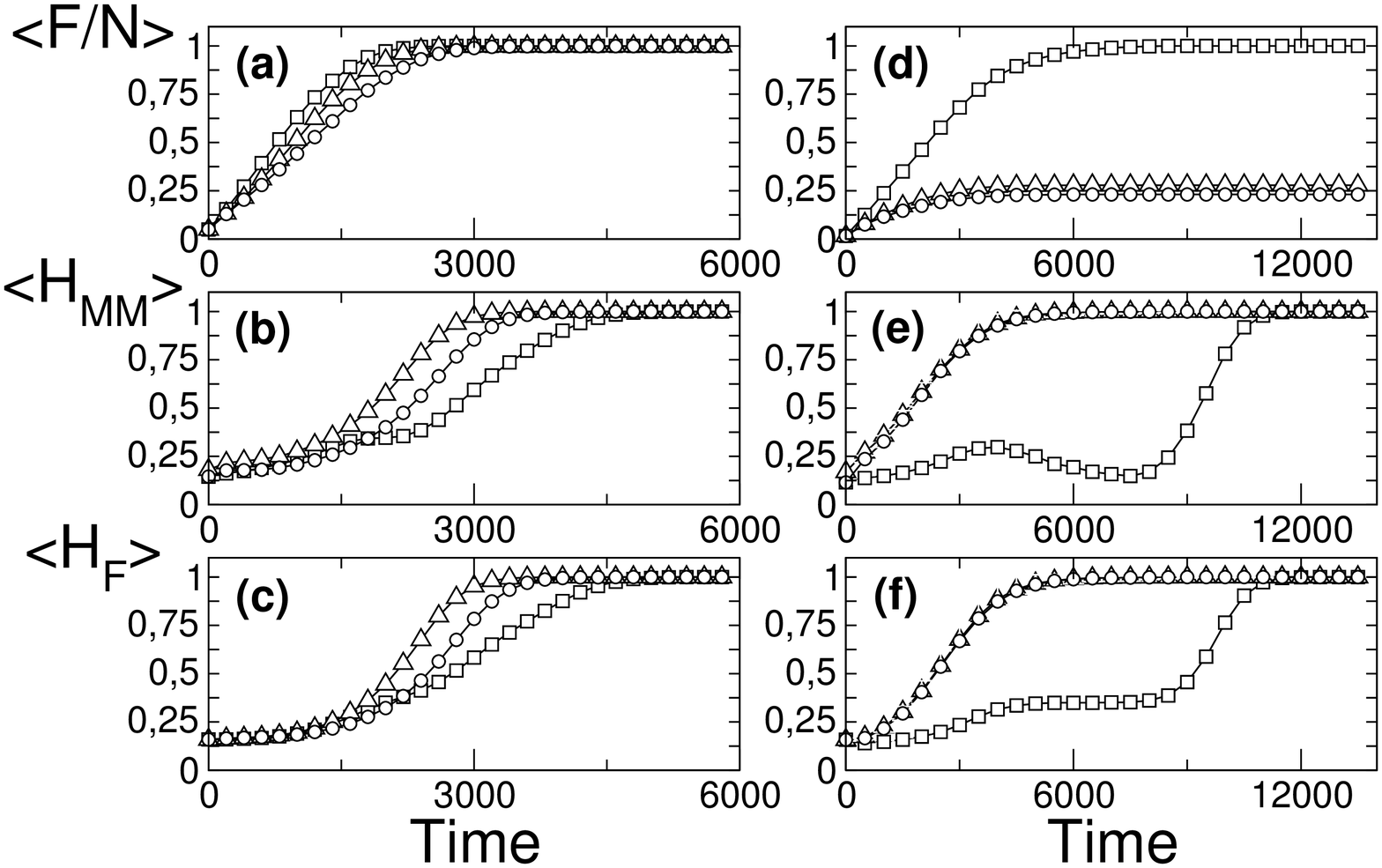}
\caption{\textbf{Dynamical 
behavior of the strategies}. 
$B=0.01$ and $Q=20$ (left panels) 
and $Q=60$ (right panels). 
Fraction of {\it Followers} 
($<F/N>$, panels (a) and (d)), 
mean homophily respect to the 
Mass Media, $<H_{MM}>$ 
(panels (b) and (e)), and mean 
homophily among {\it Followers}, 
$<H_{F}>$ (panels (c) and (f)) 
as a function of time. 
The bra-ket notation denote 
averaging over 1000 events. 
Squares denotes {\it NFS}, 
triangles {\it FS} and circles {\it FMM}.}
\label{Fig2}
\end{figure}

When we analyze the fraction of {\it Followers}
as a function of time (Fig.\ref{Fig2} panel (a)), 
we can observe that all strategies behave 
quite similar in the low $Q$ regime ($Q=20$), 
being the Non-Followers Strategy ({\it NFS}) 
the fastest and the Fixed Mass Media ({\it FMM}) 
the slowest strategy to reach the 100\% of 
{\it Followers}, as it can be seen in the table \ref{table1},
where we define $\tau$ as the time when $(F/N)$
reaches the value of 1.
However, the strategies 
produce differences among the $Followers$ 
in terms of self similarity and similarity 
with the Mass Media. If we look the behavior 
of $H_{MM}$, at the 
panel (b) of Fig.\ref{Fig2} and table \ref{table1}, 
we can see that at the time $<F/N> = 1$ 
the {\it Followers} in the {\it FS} and {\it FMM} are
closer to the Mass Media than when
we implement the {\it NFS}.
Similar behavior is found for $<H_F>$ 
(Fig.\ref{Fig2} panel(c)), showing 
that at the time of reaching consensus, 
the Mass Media which adapts to their 
followers produces a more homogeneous 
crew of them respect to the case of 
the one which adapts to 
the $Non-Followers$ agents.
The {\it Followers} attained by this strategy
form a more heterogeneous group 
until they become completely similar.\par

\begin{table}[h]
\centering
\begin{tabular}{c c c c}
Strategy & $<\tau>$ & $<H_{MM}>$ & $<H_{F}>$\\
\hline\hline
NFS & 2200 & 0.35 & 0.40\\
FS & 2400 & 0.80 & 0.70\\
FMM & 2800 & 0.80 & 0.70\\
\end{tabular}
\caption{Aproximate values of $\tau$, $H_{MM}$, and $H_{F}$ 
at the time of reaching consensus, 
in the first feature, for each strategy and $Q = 20$.
Bra-kets denote average over 1000 events.}
\label{table1}
\end{table}

In the region of large $Q$ ($Q = 60$), 
we find an unexpected non-monotonic behavior 
of $<H_{MM}>$ (Fig.\ref{Fig2} panel(e)) 
for the Non-Follower Strategy ({\it NFS}): 
At the time when $<F/N>\simeq 0.75$ 
(Fig.\ref{Fig2} panel(d)), it starts to decrease 
until the fraction of {\it Followers} become $1$, 
when it starts to increase again. 
This means that in this region 
(when the amount of {\it Non-Followers} is 
less than the 25\% of the population, 
but greater than zero), 
the similarity between the {\it Followers}
and the Mass Media is very low. 
In addition, in order to convince 
this last 25\% of the agents, 
the Mass Media takes a similar 
time interval (about 4000 time steps) 
as it took to convince 
the 75\% of the population.
What is happening in this region? 
The Mass Media tries to increase the 
probability of interaction with the 
{\it Non-Followers} changing 
its state. The {\it Non-Followers} 
can be distributed throughout all the 
lattice and they have very different 
cultural states among them. 
At the same time, when the Mass Media 
adapts to them, it departs from the 
{\it Followers}, 
which constitute the majority of the system. 
In addition, the high degree of similarity 
between the Mass Media and a small 
group of {\it Non-Followers}
doesn't favor the homogenization 
of the {\it Followers} group, as can 
be seen in Fig.(\ref{Fig2} panel(f), 
where $<H_F>$ remains constant 
during this time-lapse ($<H_F>\simeq0.35$). 
Once all the agents become {\it Followers}, 
both $<H_{MM}>$ and $<H_F>$ grow monotonically 
until they reach the value of 1 
(i.e, agents shares all feature's values with 
the Media).
In appendix \ref{sec:appendix}, we show a more detailed 
description of this behavior, 
analyzing a single event.\par
Concerning to the case of a Fixed Mass Media 
and Followers Strategy for $Q = 60$, 
both $<H_{MM}>$ and $<H_{F}>$ 
increase monotonically as it was 
observed for $Q=20$, but in this case, 
these strategies are unable to 
reach consensus. {\it FS} only reaches 
a little more than the $25\%$ 
of {\it Followers}, and {\it FMM} gets 
a percentage slightly below of that. 
The similarity among {\it Followers} 
and between the {\it Followers} 
and the Mass Media is identical for 
both strategies: they reach  
$<H_{MM}>=1$  and $<H_F>=1$ at 
almost the same time than 
they reach the largest 
amount of $Followers$ 
that they can get.\par

\subsection{Optimal combination of strategies}

The analysis performed in the previous sections tell 
us that even though the Non-Followers Strategy 
is the best one in terms of reaching consensus, 
it takes a lot of time in convincing the last fraction of agents. 
It happens because the Mass Media can change its 
state very sharply in order to maximize 
the overlap with the {\it Non-Followers}, 
which are just a few and are very different among them. 
But what would happen if the Mass Media changes 
its strategy when the homophily among {\it Followers}
stops growing (i.e., when $<F/N> \simeq 0.75$ for $Q=60$)? 
Is it possible to reach consensus 
when the second strategy is not the {\it NFS}? 
Is there an optimal balance between maximizing 
the amount of {\it Followers} and minimizing 
the time to do this? \par

\subsubsection{Temporal combinations}

\begin{figure}
\includegraphics[width = \textwidth]{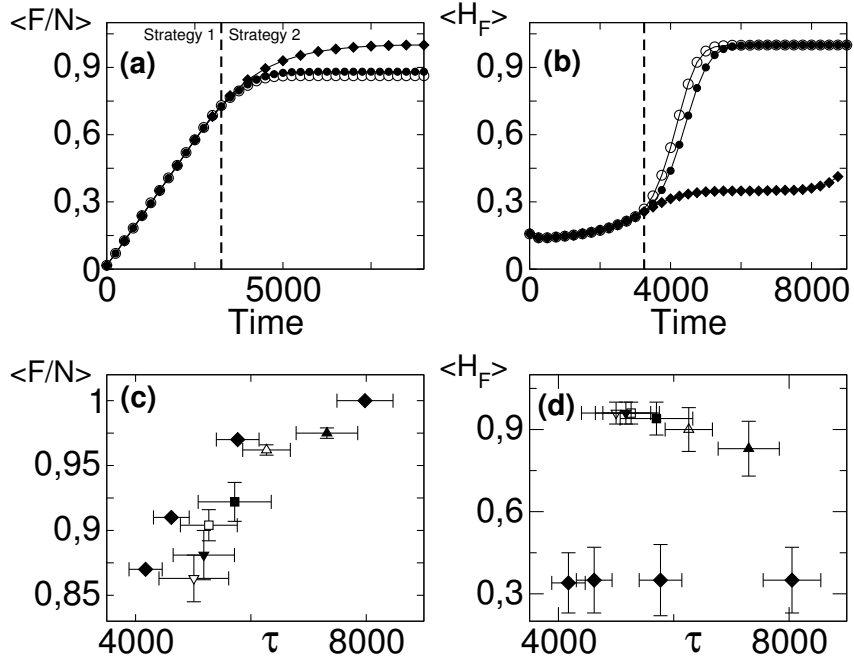}
\caption{\textbf{Combination of Strategies}. 
Panel (a), Fraction of {\it Followers} $<F/N>$, 
panel (b) homophily among {\it Followers} 
$<H_{F}>$, both as function of time for 
Non-Followers Strategy (diamonds) 
and a combination of two strategies 
starting with {\it NFS} until $(F/N)=0.75$: 
Followers Strategy ({\it FS}, empty circles) 
and Fixed Mass Media ({\it FMM}, full circles).
In all the cases, $Q=60$ and $B=0.01$. 
Panel (c), $<F/N>$, and panel (d), $<H_F>$,
as function of the time of 
reaching the asymptotic value of $<F/N>$, $\tau$.
The diamond symbol stands 
for {\it NFS}, empty symbols for {\it NFS} 
followed by {\it FS}, and full symbols 
(except diamonds) for {\it NFS} followed 
by {\it FMM}. The change of strategy is 
done at different values of $(F/N)$ 
for the {\it NFS}: $0.75$ (triangles down), 
$0.85$ (squares) 
and $0.95$ (triangles up).}
\label{Fig3}
\end{figure}

In this section we analyze how the system behaves 
when the Mass Media change its strategy at a given time.
In Fig. \ref{Fig3}, panel (a) and (b),
the Mass Media starts
with the Non-Followers Strategy ({\it NFS}) until it reaches 
the $75\%$ of {\it Followers} and then, 
it remains as a Fixed Mass Media ({\it FMM}),
or implements the Followers Strategy ({\it FS}). 
In panel (a) we can observe that,
when the combination of 
strategies is implemented, 
the Media is not able to reach the $100\%$ of 
{\it Followers}. On the other hand,
the asymptotic fraction of {\it Followers} reaches a 
value closer to $0.90$, being slightly larger 
when the second strategy is {\it FMM}, but
none of these cases can improve the {\it NFS},
which reaches that amount of {\it Followers} in 
less time. However, we can observe that the combination of strategies 
produces that $<H_{F}>$ begins to increase monotonically 
when the change is done,
in contrast to what is observed when the Media
applies a {\it pure NFS}, where it remains 
practically constant.\par

In Fig. \ref{Fig3}, panel (c) and (d),
we explore different values of $(F/N)$ in which
the Media change its strategy.
We plot $<F/N>$ and
$<H_F>$, respectively, at the time $\tau$, which is the time spent 
to reach the asymptotic value of $<F/N>$ when 
the Media applies a combination of strategies,
and we compare them with the transitory results 
obtained when only the {\it NFS} is applied. 
It is important to remark that the asymptotic fraction 
of {\it Followers} is always 1 for the {\it NFS}.
In panel (c), we can observe 
that the {\it NFS} is always the 
faster strategy to reach a given amount of 
{\it Followers}, but the combination of strategies 
produce a better homogenization when that 
value is attained (see panel (d)). 
It implies that if the Media adopts a combination of 
strategies, it relaxes the condition 
of full consensus, but the system reaches a stationary state 
(when $<H_F> = 1$ and $<F/N>$ is maximum) faster
than when a pure {\it NFS} is applied. 
However, if it wants to reach a given amount 
of {\it Followers} regardless of their 
homogenization, the {\it NFS} is the best strategy.\par

\subsubsection{Structural combinations}\label{sec:Structural}

\begin{figure}
\centering
\includegraphics[width = \textwidth]{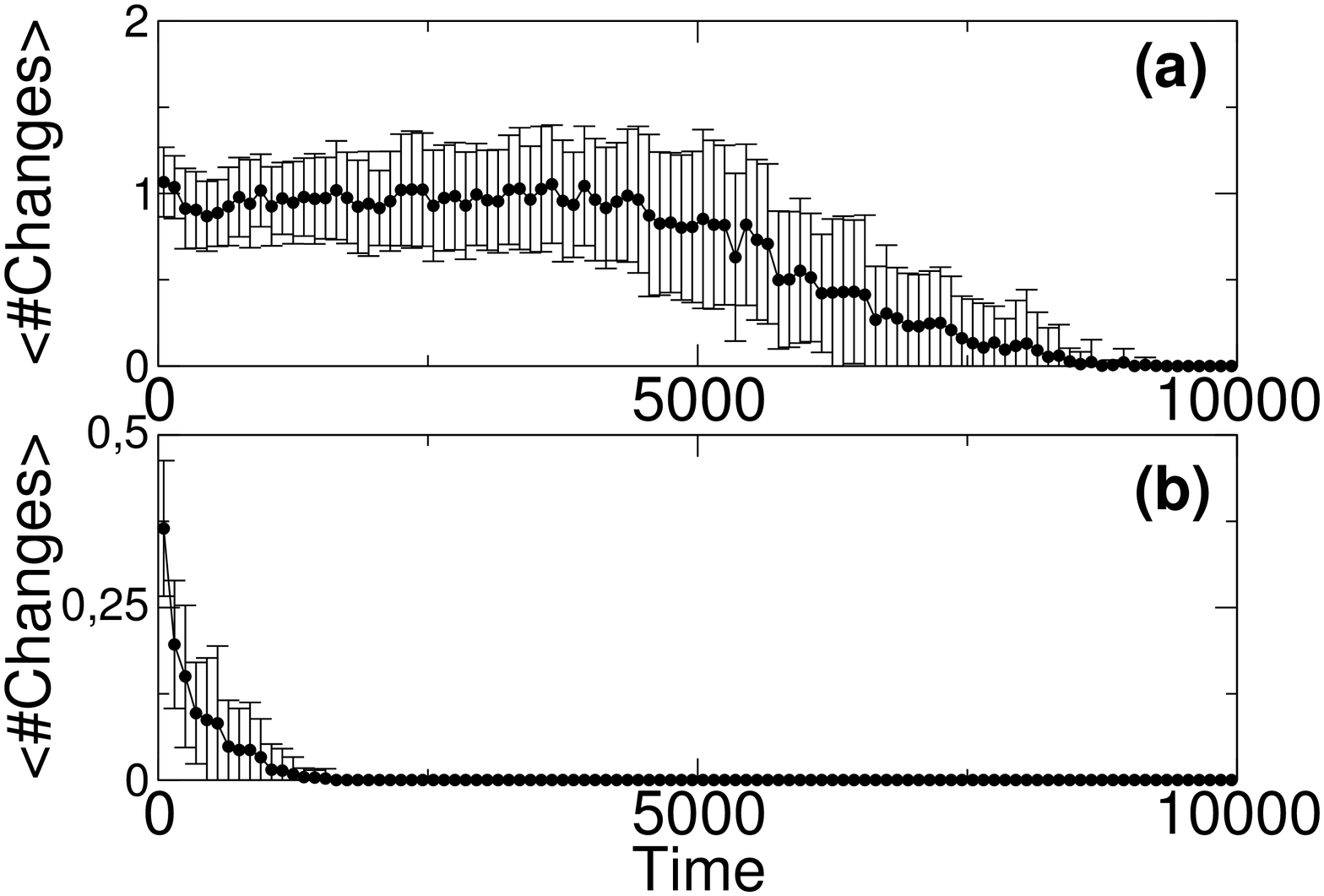}
\caption{\textbf{Average number of changes}. 
Average number of features that the Mass Media changes. 
Each point represents the mean value of changes 
over 100 time steps and this quantity is averaged over 
50 events. Panel (a) stands for NFS and panel (b) 
for FS, both with $Q=60$ and $B = 0.01$.}
\label{Fig4}
\end{figure}

In the previous sections, the Mass Media always kept fixed 
the first feature while it was able to change the values 
in the others according to the different strategies 
defined above. However when we analyze the mean 
number of changes that the Mass Media does per time step,
we found that, on average, the {\it NFS} changes just one 
feature per time step, while {\it FS} changes even less 
(and it's more similar to a Fixed Mass Media, as we 
have seen at their respective phase diagrams), as we 
can see in Fig.\ref{Fig4}.
This suggests 
that similar results can be found 
if we let the Mass Media 
change just one of the features at a time. 
This can be seen as a combination of strategies in 
the features space where one feature adapts to the 
population meanwhile the others remain fixed. 
We analyze variants of the {\it Non-Followers} strategy 
in two different cases: When the adaptive feature is 
always the same (fixed) and when it is chosen randomly 
in every time step. In all cases, the first feature 
remains constant.\par
\begin{figure}
\includegraphics[width = \textwidth]{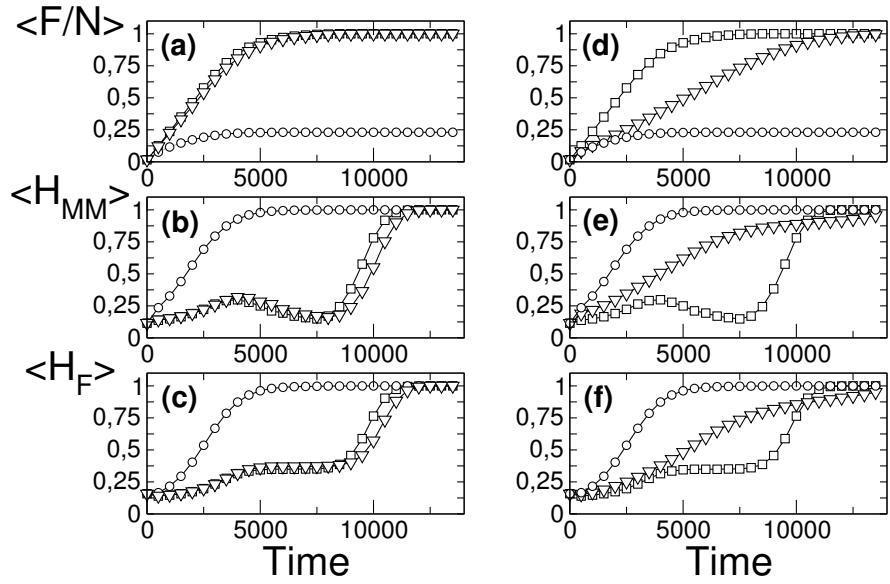}
\caption{\textbf{Combination of Strategies 2}. 
Dynamical behavior when the {\it MM} 
has only one adaptive feature for $B=0.01$ and $Q=60$. 
Fraction of {\it Followers} ($<F/N>$, panels (a) and (d)), 
mean homophily respect to the {\it MM} 
($<H_{MM}>$, panels (b) and (e)), 
and mean homophily among {\it Followers}, 
($<H_{F}>$, panels (c) and (f)) as a function of time. 
Left panels: the adaptive feature is randomly chosen
at each time step. 
Right panels: The 
adaptive feature is always the same. 
Squares denotes {\it NFS}, circles {\it FMM}, and triangles 
a Mass Media with one adaptive feature. 
Bra-kets denote average over 1000 events.}
\label{Fig5}
\end{figure}

In Figure \ref{Fig5} we plot $<F/N>$, as well as 
$<H_{MM}>$ and $<H_F>$ as 
function of time for the two cases. We analyze the system for $Q=60$ 
and $B=0.01$ and compare it with the cases of {\it NFS} 
and {\it FMM}, respectively. 
We can observe that one adaptive feature is a sufficient 
condition to reach consensus for $Q=60$ and $B=0.01$,
which is impossible if all features are fixed ({\it FMM}), 
as we have seen in Fig.\ref{Fig1}. In particular, 
if the adaptive feature is randomly chosen, 
the dynamics of the system is almost the same that in 
the case of Non-Follower Strategy ({\it NFS}). On the other hand, 
if the adaptive feature is fixed, also the system is able to 
reach consensus in regions of the parameter space 
where a Fixed Mass Media is unable to do it, but the 
convergence time is larger than the one expected for a 
full Non-Followers Strategy. On the other hand, this strategy 
favors the homogenization  of the {\it Followers} group, 
as it can be observed in the behavior of $<H_{MM}>$ 
and $<H_{F}>$ in Fig.\ref{Fig5}.\par

\section{Conclusions} \label{sec:Conclusions}

In this work we have proposed a new way to model the influence of 
a Mass Media onto a system of social agents.
Here, the Media has 
a specific purpose: To put on the agenda
a particular topic, i.e., to make people discuss a given 
topic and impose its point of view, represented by a fixed feature's value.
This way to model the Media fits better within the conflict 
theory of Media's influence (\cite{Giddens}) and within 
the works of Mccombs (\cite{Mccombs} \cite{Mccombs1972}), 
which we consider
that describe better its actual role in a society. 
In order to achieve this goal, the Media takes advantage of
the other features which are adaptives in order
to increase the probability of interaction with potencial
consumers, according to different strategies. In one of them, 
the Mass Media takes the most popular value of each 
feature among the {\it Non-Followers} which was named the 
NFS (Non-Followers Strategy). In the other one,  the Mass Media 
takes the most popular value of each feature among the 
{\it Followers} and we called it the FS (Followers Strategy). 
We compare both with the standard case where the Mass Media 
is fixed in time and then it does not follow any strategy 
at all ({\it FMM}).\par
When  the $MM$ applies the Non-Followers strategy, 
it is able to reach consensus for all values in parameter space, 
which is not the case of the Followers Strategy or 
when the MM is fixed. 
The problem with this strategy is that it takes too 
much time to reach that consensus due to the fact that 
the Mass Media ends up adopting particular agents' state 
in order to convince the last {\it Non-Follower} agents. 
This sharp changes produce that the similarity between 
the {\it Followers} and the {\it MM} decrease during this time, 
while in the other strategies it always shows an increasing 
behavior. It also produces that the {\it Followers} form 
an heterogeneous group until the last agent is convinced.\par
In order to improve the {\it NFS}, we explored different 
combinations of strategies: We have found that if the 
Mass Media combines strategies in a temporal manner, 
it can reach a large amount of $Followers$ (close to $90\%$) 
with a monotonous increase 
in their homogenization, but still a pure {\it NFS}
is the faster way to reach a given amount of {\it Followers}. 
We have also found that, 
when the combination is in the Feature Space 
(i.e., when some features are fixed and others are adaptive), 
the change of only one feature per step is a sufficient condition  
in order to reach consensus (100\% of {\it Followers}). 
Moreover, if the adaptive feature is selected at random, 
the system behaves quite similar to the case when the {\it MM} 
adopts the Non-Followers strategy. On the other hand, 
if the adaptive feature is fixed, it takes almost the double 
of time to reach the total amount of {\it Followers} 
but it produces an homogeneous group during the dynamics. 
The structural combination of strategies can be seen as a 
more economic way to have an adaptive {\it MM} that can 
reach consensus in all the parameter space.\par
This work is the first step in order to understand the 
formation of collective states when a Mass Media 
want to set the agenda and impose its point of 
view in a given feature.
Future extensions of this work should include the 
consideration of complex networks of interaction, and
the presence of two or more Media in a competitive
context.\par
\appendix

\section{Microscopic description of {\it NF} strategy} \label{sec:appendix}

\begin{figure}
\centering
\includegraphics[scale = 0.50]{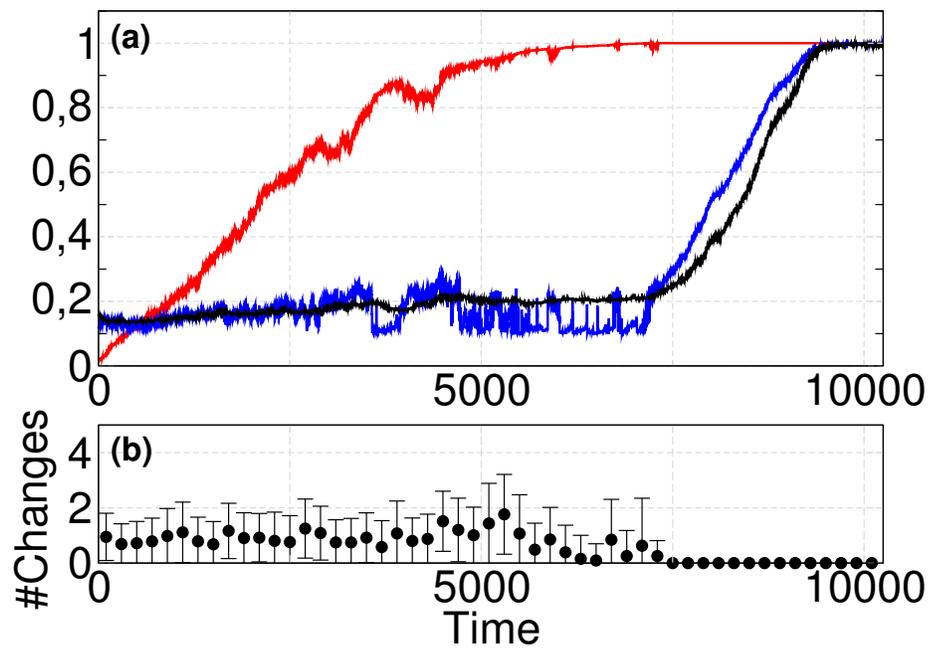}
\caption{\textbf{(Color online) Dynamical behaviour of 
single event of 
the Non-Follower Strategy}. $B=0.01$ and $Q=60$. 
Panel (a): Fraction of {\it Followers} (($F/N$), red line), 
mean homophily between {\it Followers} 
and the {\it MM} ($H_{MM}$, blue), and mean homophily 
among $Followers$, ($H_{F}$, black) as a function of time. 
Panel (b): Average change of the {\it MM} as 
function of time.}
\label{Fig_appendix}
\end{figure}

In this appendix we describe the peculiar behavior of the {\it NFS}: 
It is very good to collect {\it Followers} quickly but it loses 
a lot of time in trying to convince the last fraction of the population. 
We can see in detail this behavior and understand the 
counter-intuitive dynamics of this strategy 
by analyzing a single representative dynamical event.
In Figure \ref{Fig_appendix}, panel (a), 
we compute $(F/N)$, $H_{MM}$, and $H_{F}$, as function of time 
for a single event with $Q = 60$ and $B=0.01$. 
In panel (b), we compute the average changes that 
the {\it MM} does as function of time. 
The average is over 100 time steps.\par
\begin{figure}
\centering
\includegraphics[scale = 0.80]{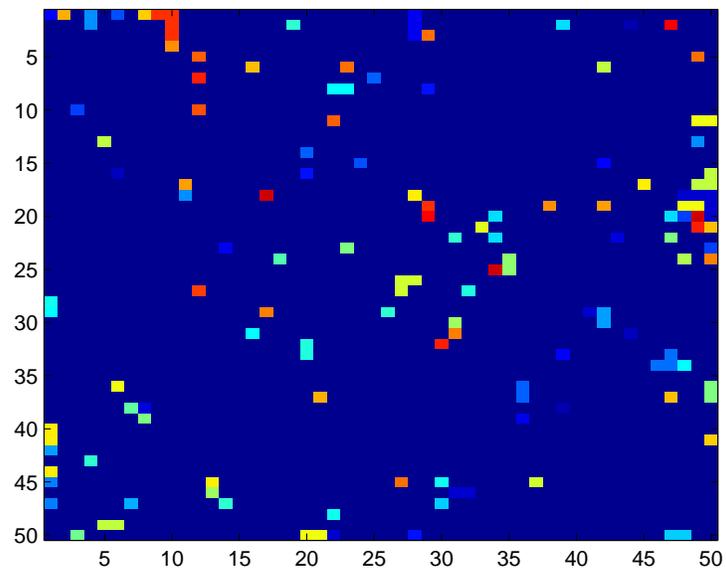}
\caption{\textbf{(Color online) Snapshot}. It corresponds to $t = 5000$, 
while the colors denote different values in the first 
component of the agents' states. 
The most abundant color represents the {\it Followers}.}
\label{Fig_appendix2}
\end{figure}

If we see first $H_{MM}$, we notice that it doesn't 
present the smooth decreasing behavior observed 
in Fig.\ref{Fig2} panel (e), when $<H_{MM}>$ represented 
the average over many events. In a single event, we can see 
that this decreasing region is replaced by a ``noisy" region, 
which extends from $t \simeq 4000$ to $t \simeq 7500$ 
in this particular event. 
In this region, less than the $25\%$ of the agents are 
{\it Non-Followers}, and it's naturally to think that 
they are distributed over the lattice, i.e. they don't 
belong from the same cluster, as it's suggested 
by the snapshot of Fig.\ref{Fig_appendix2}. 
Then, when the {\it MM} looks for the most abundant 
$Q$ for each feature, it finds that this one is not 
well defined, i.e. there is no value of $Q$ for a given 
feature shared by a great number of {\it Non-Followers}. 
Instead of that, the {\it MM} finds a lot of $Q$ values 
with the same frequency (this is also due to the fact that $Q=60$, 
which decrease the probability that two agents share the same 
value of $Q$ in a given feature). 
Therefore, the {\it MM} adopts a state which can be a 
mixture of different agents or, in the extreme case, 
a state equal to a single agent, except in the first feature.
The latter situation becomes more likely when the number 
of {\it Non-Followers} tends to zero, 
being accurate when there is only one agent without convincing.
By adopting a state similar to a very few agents, 
the {\it MM} departs from the great number of 
{\it Followers} present in the system, resulting 
in a decrease of $H_{MM}$, showing the ``noisy" behavior 
with a lower bound in $0.1$, which represents the minimum 
homophily between the {\it MM} and a {\it Follower}, 
which by definition, shares at least the first feature.
At the same time, the sharp changes of the {\it MM} does 
not help to homogenize the {\it Followers} agents, 
which ``become confused" by the {\it MM} 
messages and maintain constant a very low similarity 
among them ($H_F$).\par
The ``noisy" region finishes at $t \simeq 7500$, when all 
agents become {\it Followers}. As it can be seen in 
Figure \ref{Fig_appendix} panel (b), from this moment 
the {\it MM} doesn't change, adopting the state of the 
last {\it Non-Follower} convinced. So, both $H_{MM}$ 
and $H_{F}$ grow monotonically to a value of $1.0$, 
when all agents finish to adopt the Mass Media's state. 
Before it has stopped to change, the average number 
of changes is about 1 change per time step, 
as it was observed in section
\ref{sec:Structural}.\par

\newpage

\bibliography{AxelrodMM}

\newpage

\end{document}